\documentclass[twocolumn,showpacs,preprintnumbers,aps,prl,10pt,nofootinbib,superscriptaddress]{revtex4}

\usepackage[english]{babel} 	
\usepackage{graphics} 
\usepackage[pdftex]{graphicx}
\usepackage{amssymb} 			
\usepackage{amsmath} 			
\usepackage{booktabs}
\usepackage{tabularx}
\usepackage{rotating}
\usepackage{braket}

\usepackage{float}
\usepackage{url}	
\usepackage{bm}

\graphicspath{{./figures/}}

\begin{document}

\title{Emergence of clusters: Halos, Efimov states, and experimental signals}




\author{D. Hove}
\affiliation{Department of Physics and Astronomy, Aarhus University, DK-8000 Aarhus C, Denmark}

\author{E. Garrido}
\affiliation{Instituto de Estructura de la Materia, IEM-CSIC, Serrano 123, E-28006 Madrid, Spain}

\author{P. Sarriguren}
\affiliation{Instituto de Estructura de la Materia, IEM-CSIC, Serrano 123, E-28006 Madrid, Spain}

\author{D.V. Fedorov}
\affiliation{Department of Physics and Astronomy, Aarhus University, DK-8000 Aarhus C, Denmark}

\author{H.O.U. Fynbo}
\affiliation{Department of Physics and Astronomy, Aarhus University, DK-8000 Aarhus C, Denmark}

\author{A.S. Jensen}
\affiliation{Department of Physics and Astronomy, Aarhus University, DK-8000 Aarhus C, Denmark}

\author{N.T. Zinner}
\affiliation{Department of Physics and Astronomy, Aarhus University, DK-8000 Aarhus C, Denmark}

\begin{abstract}
We investigate emergence of halos and Efimov states in nuclei by use of a newly designed model which combines self-consistent mean-field and three-body descriptions. Recent interest in neutron heavy calcium isotopes makes $^{72}$Ca ($^{70}$Ca+n+n) an ideal realistic candidate on the neutron dripline, and we use it as a representative example that illustrates our broadly applicable conclusions. By smooth variation of the interactions we simulate the crossover from well-bound systems to structures beyond the threshold of binding, and find that halo-configurations emerge from the mean-field structure for three-body binding energy less than $\sim 100$~keV. Strong evidence is provided that Efimov states cannot exist in nuclei. The structure that bears the most resemblance to an Efimov state is a giant halo extending beyond the neutron-core scattering length. We show that the observable large-distance decay properties of the wave function can differ substantially from the bulk part at short distances, and that this evolution can be traced with our combination of few- and many-body formalisms. This connection is vital for interpretation of measurements such as those where an initial state is populated in a reaction or by a beta-decay.
\end{abstract}


\maketitle

\paragraph{Introduction.}

Constituent particles of $N$-body nuclear structures are usually approximated as point-like nucleons where the intrinsic degrees of freedom are frozen.  The simplest structure beyond the deuteron is the uncorrelated average structure from the mean-field assumption of antisymmetrized product wave functions \cite{vau72,dob16,nik11,vau73,ben03}. Often correlations are decisive and require special treatment \cite{bar13,epe09,car15,her16,ono92,fel97,hag10,tan78}.

Nuclear structures can vary from spherical mean-field properties, over
collective deformations and a variety of other correlations, to bound nuclear
clusters each in (almost) inert subsystems
\cite{epe09,lei13,hag14,lie16}. Very crudely, we can say that
nuclei around beta-stability are fairly well described by
self-consistent mean-field calculations while approach to the
nucleon driplines produces two- and three-body halo structures
\cite{jen00,fre12}.  For excited states at energies close to
threshold for cluster separation, the corresponding clusterization
is strongly favored \cite{erl12}.

If the clusters, in addition to weak binding, have a sufficiently large $s$-wave scattering length, the formation of the much coveted Efimov states is theoretically possible \cite{nie01,fed93,jen03,gar06,efi70}. Many different nuclei have been proposed as viable candidates for the formation of nuclear Efimov states \cite{jen03,maz00,maz97}. Most recently the neutron rich end of the calcium isotope chain was suggested as a likely candidate for Efimov physics \cite{hag13}.

For heavy systems traditional clusterized few-body techniques seem problematic, as they tend to neglect the internal structure of the clusters. In a newly designed model the core with many nucleons is treated in a mean-field approximation and the valence particles with few-body techniques \cite{hov17}. This model allows us to investigate emergence of halo and Efimov states close to the neutron dripline, using $^{72}$Ca as the test case.  This choice is very well suited for several reasons: mean-field computations find that $^{70}$Ca is spherical closed shell dripline nucleus, $^{72}$Ca is mean-field unbound, but a stable borromean nucleus in three-body calculations. The $s$-state near the Fermi level provides optimal conditions for pronounced halo structures and possibly Efimov states.

The purpose of the present letter is three-fold; (i) to discuss
the emergence of halos, (ii) examine the possible existence of Efimov-states for realistic
effective interactions, and (iii) make a connection between short- and
large-distance quantum solutions.

\paragraph{Model ingredients.}

Our starting point is an $A+2$ nucleon system viewed as a core with mass number $A$ and two valence nucleons. We assume two- and three-body interactions, $V_{ij}$ and $V_{ijk}$, acting between nucleons. Our general Hamiltonian is 
\begin{align}
H=\sum_{i=1}^{A+2} T_i - T_{cm} + \sum_{i<j}^{A+2} V_{ij} + \sum_{i<j<k}^{A} V_{ijk} + V_{3b}, 
\label{eq:fullHam}
\end{align} 
where $T_i$ and $T_{cm}$ are kinetic energy operators for the $i'$th
nucleon and the total $A+2$ system respectively, and $V_{3b}$ is the three-body interaction between the valence neutrons and the core particles. 
Central to our method is the choice of Hilbert space and the corresponding interactions. We shall use Skyrme forces and wave functions of the form
\begin{align}
\Psi= \mathcal{A} \left( \Phi_c(\bm{r}_1,\cdots,\bm{r}_A) \Phi_v(\bm{r}_{v_1},\bm{r}_{v_2}) \right),
\label{eq3}
\end{align}
where $\Phi_c$ is the Slater determinant of core nucleons with spin and space
coordinates, $(\bm{r}_1,\cdots,\bm{r}_A)$, $\Phi_v $ is the three-body
wave function with nucleon-core relative coordinates,
$(\bm{r}_{v_1},\bm{r}_{v_2})$, and $\mathcal{A}$ is
anti-symmetrization operator.

Equation~(\ref{eq3}) clearly exhibits how we combine mean-field treatment of the core nucleons with few-body treatment of the two relative degrees of freedom.  We emphasize that the effective interactions determining $\Phi_c$ depends on $\Phi_v$ and vice versa.  The equations for the lowest energy solution is obtained by varying the wave function, $\Psi$, over the allowed Hilbert space, that is
\begin{align} 
\frac{\delta \langle \Psi|H|\Psi \rangle}{\delta\Phi_c^*} 
= \frac{\delta \langle \Psi|H|\Psi \rangle}{\delta\Phi_v^*} 
= 0\;,
\label{varyeq}
\end{align}
where both $\Phi_c$ and $\Phi_v$ must be normalized throughout the variation. The resulting coupled Hartree-Fock and three-body equations determine the two wave functions and subsequently the energy.

The three-body equation is solved using the hyperspherical adiabatic expansion method \cite{nie01}. We exclude the core-occupied Pauli forbidden states either by removal of the associated hyperspheric states or by using phase equivalent potentials \cite{gar99}. The final step is solving a coupled set of hyper-radial equations where the simplest, but still realistic version of the hyperradial equations for extreme halos, can be written as
\begin{align} 
\left(
-\frac{\partial^2}{\partial \rho^2} - \frac{\xi^2(\rho) + 1/4}{\rho^2} +  v_{short}(\rho) - \tilde{\epsilon} 
\right) f(\rho)  
= 0  \;,
\label{radial}
\end{align}
where $\rho$ is the hyperradius \cite{nie01}, $f(\rho)$ is the dominating radial
wave function, and $v_{short}(\rho)$ and $\tilde{\epsilon}$ are reduced
short-range interaction and three-body energy.  The parameter,
$\xi(\rho)$, is obtained from solving the angular part of the
three-body equations.  For at least very large $s$-wave scattering
lengths $\xi$ is constant over a substantial $\rho$-interval, where
Efimov states might exist.  Energies and mean square radii of
neighboring solutions are related through the scaling, $s^2 =
\exp(2 \pi/\xi)$.  The $\rho$-dependence of this parameter is then of
crucial importance for occurrence of halos and existence of Efimov
states.

\paragraph{Why $^{72}$Ca ($^{70}$Ca+n+n).}

The ideal system of choice is a spherical, closed shell, dripline nucleus. Our mean-field calculations with the SLy4 parametrization \cite{cha98} predict that $^{70}$Ca is bound while $^{72}$Ca is unbound. This is in general agreement with other mean-field calculations of various types \cite{mad05,erl12}.  However, $^{72}$Ca ($^{70}$Ca+n+n) is
bound in a three-body calculation where the two-body subsystems are unbound.  Thus, $^{72}$Ca is a borromean nucleus at the
neutron dripline within our model and a perfect test case for the method where the core
is treated by the mean-field method.

Seeing that Skyrme interaction parameters are mainly fitted to the properties of experimentally well-known nuclei, it is not surprising that the predictions of the various Skyrme parametrizations differ when dealing with exotic nuclei. This is particularly true for the prediction of the neutron-dripline isotopes that may change significantly depending on the interaction. For Ca isotopes one finds neutron driplines ranging from A=68-76 \cite{erl12}.

A coupled cluster model instead predicts the neutron
dripline to be at $^{60}$Ca \cite{hag13}.  The difference is
apparently that the $g_{9/2}$-level is much higher in these calculations, and the $s_{1/2}$-level is the first unoccupied level in
$^{60}$Ca. 

The three-body part with our method would be similar for both
$^{62}$Ca and $^{72}$Ca, deviating only by the piece of the
neutron-core interaction arising from the filled $g_{9/2}$-levels.  The
essential $s_{1/2}$-level is unoccupied in both models.
Therefore, the discussion about appearance of halos and Efimov states
would be the same.  The proximity of the dripline is the decisive
property, since approximate neutron-core decoupling, along with weak binding and large spatial extension, is then most likely.  We
shall vary the interactions and investigate the emergence of halo
structures and the possible existence of Efimov states. Specifically, the global Skyrme parameters, usually denoted $t_{\{0-3\}}$, are scaled, such that $t_i \rightarrow S t_i$.

\paragraph{Basic two-body properties.}

We want to study how cluster structure emerges when neutrons are added and the dripline
approached. To be practical we simulate the dripline approach by continuous scaling of the interaction for the same number of nucleons. The globally determined Skyrme parameters have in any case to be fine-tuned to be appropriate for specific nuclear regions.

\begin{figure}[t]
\centering
\includegraphics[width=1\linewidth]{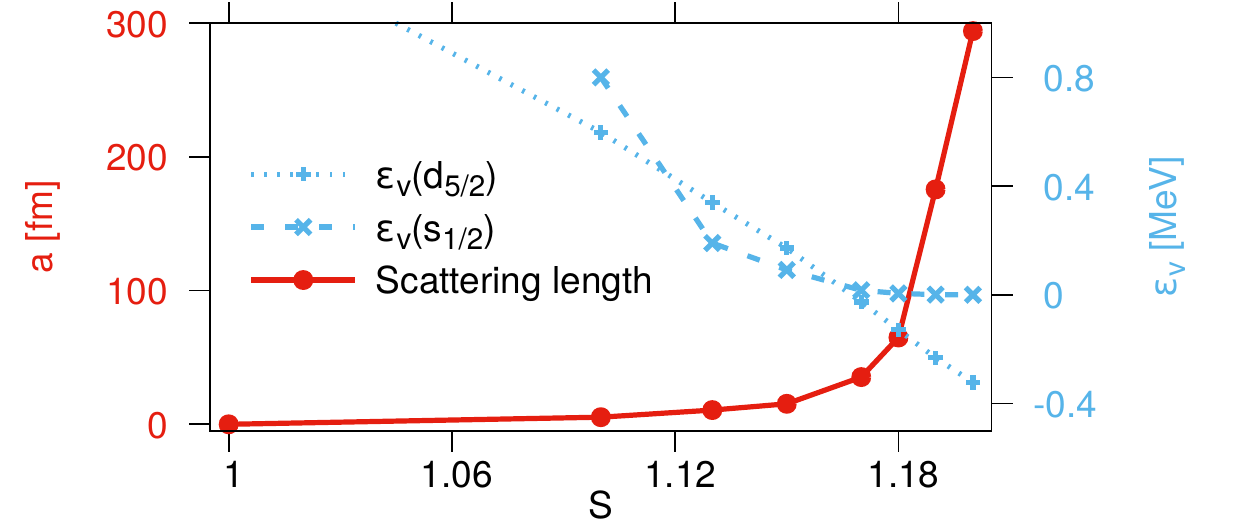}
\vspace*{-6mm}
\caption{The $s$-wave $n$-core scattering length, $a$, (red, solid) as function of the scaling of the Skyrme-parameters, $S$. The virtual energy of $s_{1/2}$ (blue, dashed) is calculated as $\hbar^2/(2 \mu a^2)$, where $\mu$ is the reduced mass, while the energy of the $d_{5/2}$ (blue, dotted) is obtained from the corresponding invariant mass spectrum, when it is unbound, and from a two-body calculation, when bound.} \label{fig4:scat}
\end{figure}

The crucial quantity for appearance of halos and Efimov states is the
$n$-core $s$-wave interaction which must have a bound, virtual, or
resonance state close to zero binding.  The present mean-field plus
three-body method folds this two-body interaction
with the calculated core structure.  The resulting neutron-core
scattering length, $a$, is shown for $^{70}\text{Ca}+n$ in
fig.~\ref{fig4:scat} as function of the overall scaling of the Skyrme
parameters.  The threshold for $s$-wave binding, $|a| = \infty$, is
reached when the Skyrme parameters are increased by a
factor of $S = 1.201$.  The estimated energy of the related
virtual $s$-state is also shown as a measure of the distance from the threshold.
The significant variation in the scattering length takes place when the virtual 
energy is less than about $300$~keV.

Scaling the Skyrme parameters shifts all single-particle energies, and
in particular the $d_{5/2}$ neutron level located immediately above
the Fermi surface of $^{70}$Ca. This state becomes
bound for $S = 1.17$ as seen in fig.~\ref{fig4:scat}.

\paragraph{Two-neutron halos.}

Halo structures are characterized by very small binding energy and large radius which can be
achieved most easily with high neutron occupancy of valence $s$-waves.
Scaling the Skyrme interaction or varying the three-body
potential can both change the structure from confined to spatially
extended two-neutron halo configuration.  One complication in the
present example is that the short-distance ground state configuration
is dominated by $n$-core $d$-waves. Only the large-distance tail outside the
$d$-wave centrifugal barrier is then expected to be $s$-wave
dominated.  

\begin{figure}[t]
\centering
\includegraphics[width=1\linewidth]{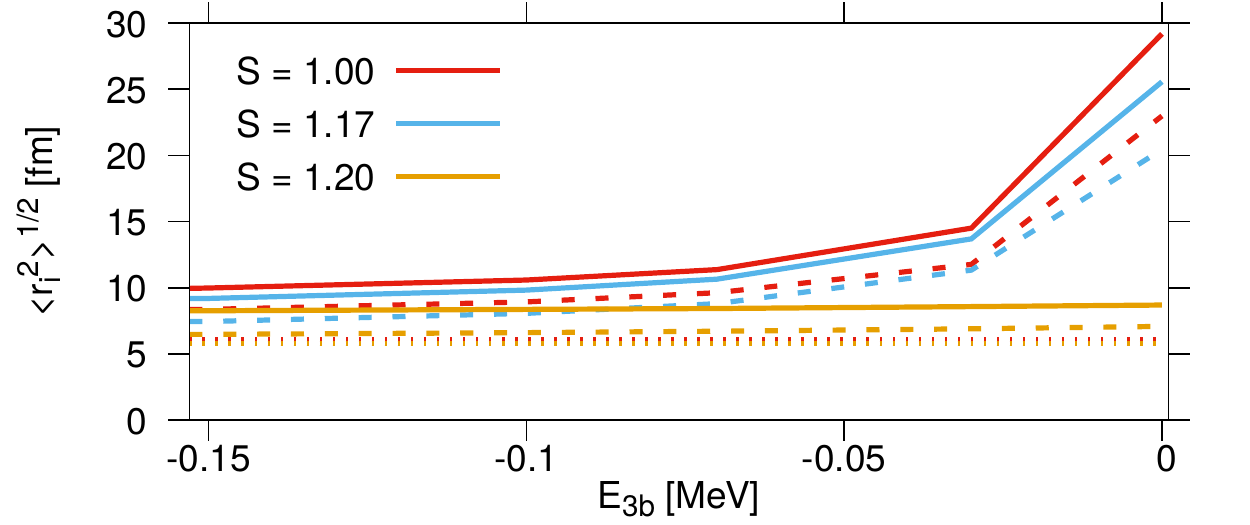}
\vspace*{-6mm}
\caption{ Rms distances as functions of the three-body energy, $E_{3b}$. The two-body systems distances are valence $n$-$n$ $\braket{r_{n,n}^2}^{1/2}$ (solid), valence $n$-core $\braket{r_{c,n}^2}^{1/2}$ (dashed), and core $n$-$n$ $\braket{r_{c,c}^2}^{1/2}$ (dotted) radius for three different $S$-values. }
\label{fig3:radii}
\end{figure}

The simplest descriptive structure properties, related to spatial extension, are average distances. In fig.~\ref{fig3:radii} we exhibit rms radii as function of the three-body energy below two-body threshold, obtained by adjusting the three-body potential, for three different values of $S$.

The average distance between valence particles (solid curves) for relatively large binding energy is substantially larger than the corresponding average distances between the nucleons inside the core (dotted curves). This difference reflects that the two neutrons are located at the top of the Fermi sea or at the surface of the nucleus. The separation of the valence neutrons is larger than the core radius, $\sim 5$~fm, indicating a tendency to be located on opposite sides of the core. Decreasing the binding energy towards zero leaves the $n$-$n$ distances inside the core unchanged while the valence neutrons move away. Halo configurations emerge clearly for the two borromean cases (red and blue curves) around a two-neutron binding of $100$~keV where the valence neutron distance increases rather abruptly.

The third case with $S = 1.20$ in fig.~\ref{fig3:radii} would produce
a bound two-body $d_{5/2}$-state with an energy approached by the
adiabatic potential at large distance.  Decreasing the binding energy
towards zero with respect to this threshold populates $d$-waves and the
rms radii remain finite as the $d$-waves prohibit
spatially extended halo configurations. This fact
shows that the appearance of the halo structure when approaching the dripline 
is not an inherent characteristic of the method, but a phenomenon that requires 
some particular conditions accounted for by the method itself.

\begin{figure}
\centering
\includegraphics[width=1\linewidth]{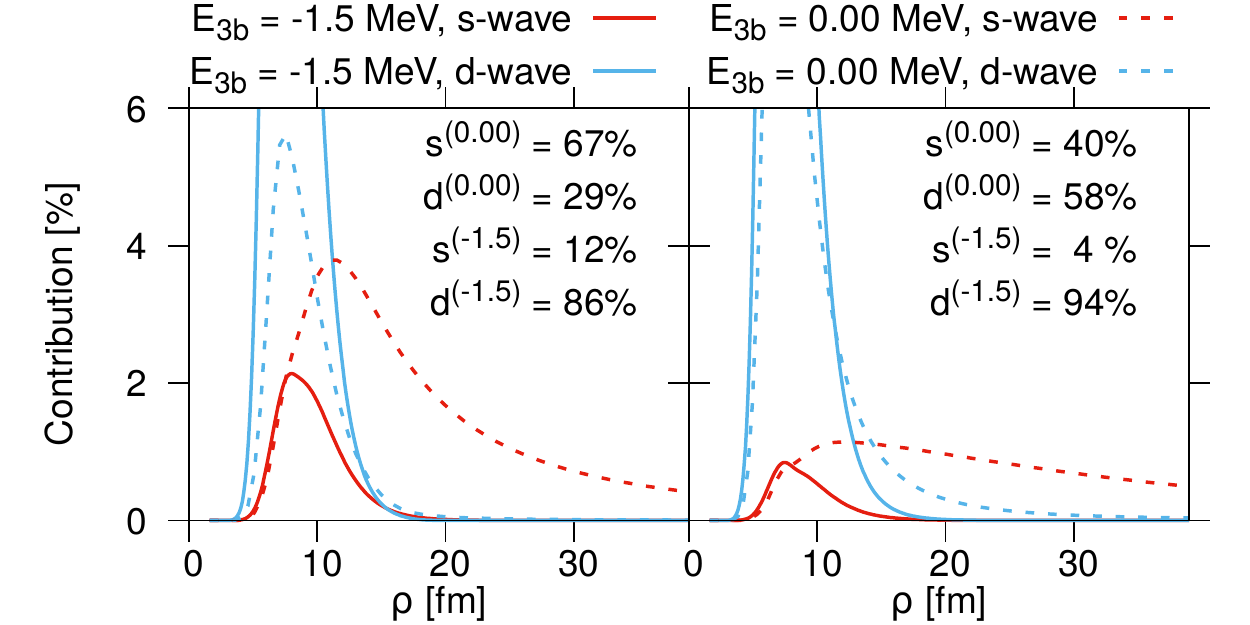}
\vspace*{-6mm}
\caption{The partial wave contributions from $s$- (red) and $d$-waves (blue) for $n$-core as functions of hyperradius for $S=1.0$ (left), and $S=1.17$ (right). Full and dashed lines indicate a three-body energy of $-1.5$~MeV and $0.00$~MeV, respectively. The total contribution after integration over $\rho$ is also given. \label{fig5:partial}}
\end{figure}

The variation of $s$- and $d$-wave contributions can be
seen in fig.~\ref{fig5:partial} as functions of $\rho$, with $S=1.0$ (left) and $S=1.17$ (right).  The three-body interaction is adjusted to give the
quoted binding energies.  The partial wave decomposition in
fig.~\ref{fig5:partial} reveals that the $s$-wave tails (red curves) 
strongly increase with decreasing binding energy.  The $s$-wave content is
determined by the contributing lowest-lying adiabatic potentials,
which in turn inherit the character of the lowest-lying $n$-core
single-particle structure. As a result the $s$-wave contribution is
smaller for the scaled Skyrme interaction where the $d$-wave energy is
closer to zero.  Thus, the large-distance tails of the wave functions may change characteristics by fine-tuning the Skyrme interaction.

The traditional two-fold criteria for the formation of halos is a very weakly bound two-body $s$- (or possibly $p$-) state. From fig.~\ref{fig3:radii} we see how a weak binding, in the present nuclear case less than about $100$~keV, leads to an extended state. However, the second part of the condition is seen to be more complicated from fig.~\ref{fig5:partial}, where even a state with a dominating $d$-wave can lead to an extended halo structure provided a slight $s$-wave tail exists.

The above discussion about halo formation in ground states must be
supplemented by the possibility of halos as excited states.  By
definition these states are less stable and often very fragile.  The
decisive properties are still small binding energy and dominating
$s$-wave structure.  States built on the second adiabatic potential
are then tempting to study, since they are dominated by $s$-waves.

\begin{figure}
\centering
\includegraphics[width=1\linewidth]{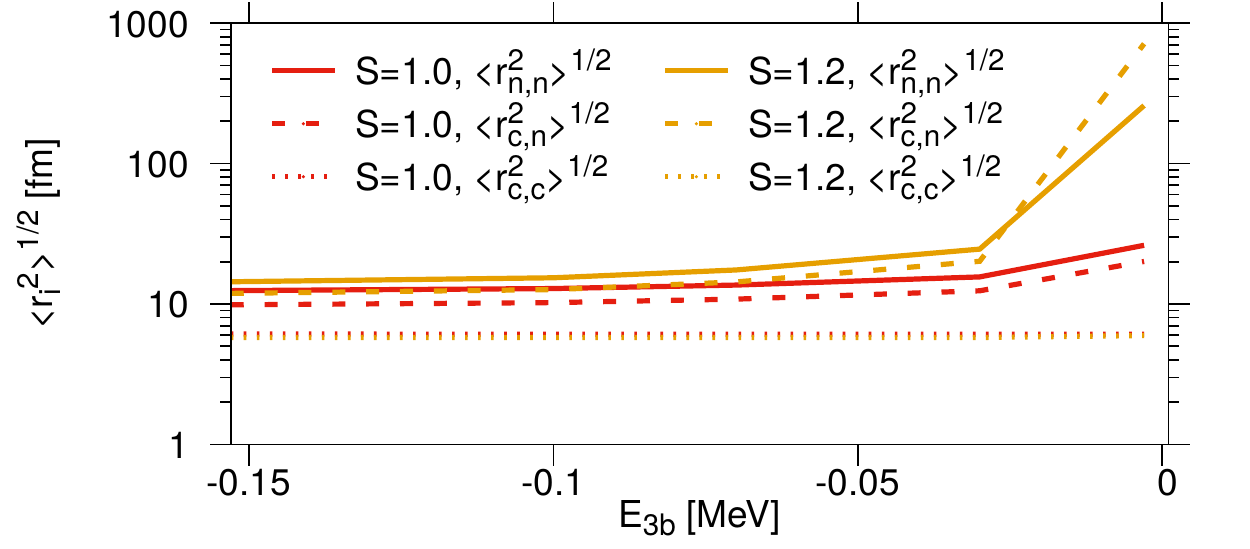}
\vspace*{-6mm}
\caption{The same as fig.~\ref{fig3:radii}, on a logarithmic scale, for the first excited three-body state with $S=1.0$ (red) or $S=1.20$ (orange).}
\label{fig3:excradii}
\end{figure}

Unfortunately, the $d_{5/2}$-state is then bound and
excited three-body states built on the second adiabatic potential
are discrete states on top of the two-body continuum from the
first adiabatic potential.  The states are therefore much more
difficult to calculate, but still well defined and possible to
obtain as demonstrated in fig.~\ref{fig3:excradii}. There is an overall similarity to the ground state results in fig.~\ref{fig3:radii}, except that all radii are now larger.  There
is, however, a significant qualitative difference for $S=1.20$.  Instead of remaining finite
at threshold the valence related radii both increase enormously over
the last $50$~keV.

\paragraph{Efimov states.}

This fast increase in size discussed in the previous section is the build up of a
halo, or theoretically the beginning of an Efimov state. In contrast to halos, for an
Efimov state to appear it is not sufficient to have a small two-neutron 
separation energy. A very large $s$-wave scattering length, $a$, is also required, which is mathematically known to produce an $1/\rho^2$ potential of the form in
Eq.~(\ref{radial}) with $\xi^2 > 0$ and constant for distances much smaller than
$a$. The convergence is extremely slow and only reached when $\rho$ is
several orders of magnitude smaller than $a$, but still much larger than the
light-light scattering length, which here is roughly 20~fm. This can only be achieved when $a$ is
larger than $\sim 1000 \times 20 = 20000$~fm, which is very unrealistic for nuclei.

But let us assume the precursor of the Efimov series is obtained with a modest
size of at least the range of the core potential, $r_0$.  The Efimov
scaling, $s$, then predicts the radius of the next state to be $s$
times larger.  We find $s \simeq 15$ \cite{fre12,jen03,bra06}
from an assumption of three simultaneously large scattering lengths.
This assumption is inherently true for zero-range models.  However, this
assumes that the neutron-neutron scattering length is much larger than
both $r_0$ \cite{fre12,bra06}, and the size, $sr_0$, of the first
Efimov state.  Since the neutron-neutron scattering length is around $20$~fm this state extends into the region of only two large scattering
lengths. In this case, the scaling is determined by (see Ref.~\cite{jen03})
\begin{align}
\xi \cosh(\xi \pi/2) \sin(2 \phi) = 2 \sinh( \xi(i/2 - \phi)),
\label{eq:xi}
\end{align}
with $\phi = \arctan (\sqrt{(m_c(m_c+2m_n)}/m_n)$, where $m_c$ and $m_n$ are the core and neutron masses. For $^{72}\text{Ca}$ this results in $\xi \simeq 0.01035$ which leads to an absurdly large scaling of $s \sim 10^{131}$. 

Clearly there is no room for an Efimov state with these scaling
properties, but this does not mean that no excited bound state can occur
for a moderate scattering length of $300$~fm.  The structure of the
largest state in fig.~\ref{fig3:excradii} is a ``two-body'' halo state
in the hyperspherical coordinate describing the $s$-wave dominated
adiabatic potential.  Clearly it is located at a distance outside the
scattering length, that is in the ``classically forbidden'' region in
these coordinates.  Consequently, this state does not scale as an
Efimov state, even though it still can be characterized as an enormous
halo state.  This discussion is very general and applies to all
nuclei, including the sometimes proposed "nuclear Efimov candidates" such as $^{62}$Ca \cite{hag13} where $\xi \simeq 0.01205$ and $s \sim 10^{113}$, and also $^{11}$Li \cite{jen03} where $\xi \simeq 0.07382$ results in a scaling of $s \sim 10^{16}$, which is less extreme, but still enormous.

It is not surprising that Efimov states cannot appear, with a reasonable scaling, given that the system necessarily consists of one heavy and two light clusters with a relative small scattering length. A series expansion of Eq.~(\ref{eq:xi}) leads to
\begin{align}
|\xi| \approx \frac{4 m_{\text{light}}}{ \sqrt{3} \pi m_{\text{heavy}}}.
\end{align}
The exponential dependence on mass ratio effectively excludes the possibility of nuclear Efimov states.

\paragraph{Measurement interpretation.}

The final state, single-particle energy distribution is to a large extent dictated by the scattering lengths \cite{gar06}. The single-particle energy distributions are also the simplest non-trivial observables to investigate, where a pertinent question is which information can be extracted from such a measurement.  This can also be formulated by asking which path did the
particles take before reaching the detectors, and perhaps even what
was the decaying bulk structure located at small distances.
This formulation is usually not meaningful in quantum mechanics where
we have to be satisfied with a probability description for each path
and initial state.  However, the calculated wave function relates the
observable large- and short-distance bulk properties.

\begin{figure}
\centering
\includegraphics[width=1\linewidth]{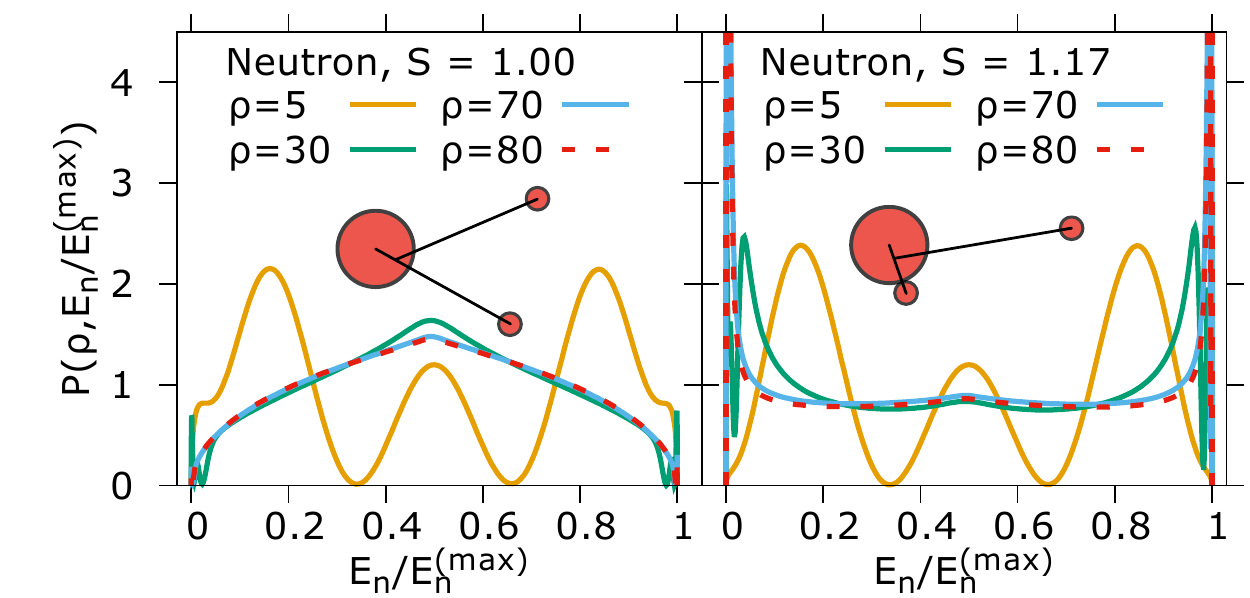}
\vspace*{-6mm}
\caption{The single-particle energy distributions for the neutron at $\rho = 5, \, 30, \, 70$, and $80$~fm, for $S=1.0$ (left) and $S=1.17$ (right). Schematic illustrations of the large distance configurations are also included. }
\label{fig:energyDist}
\end{figure}

In fig.~\ref{fig:energyDist} the single-particle energy distributions \cite{gar07,fyn09} are seen for a halo state with a total three-body energy at or very slightly above zero. This is shown for $S=1.0$, corresponding to the last point on the red curve in fig.~\ref{fig3:radii}, and for $S=1.17$, corresponding to the last point on the blue curve. In both cases the probability at around $\rho = 5$~fm shows an oscillating distribution
with three peaks.  As $\rho$ increases these oscillations are smoothed
out and the observable energy distributions appear at large distances. For $S=1.0$ the neutrons have an extended flat symmetric peak at half the maximum
energy.  This corresponds to a configuration with both the two neutrons moving away
from the core.  The pattern is typical for decay directly into the continuum without an intermediate
stepping stone as in sequential decay.

For $S=1.17$ the initial configuration is very similar, but it evolves radically differently. From the three peak structure it transitions into a structure with one of the neutrons close to the core, while the other moves away, as encouraged by a $d_{5/2}$-level close to zero energy. This structure is then a signal of dominating two-body attractions between the unequal particles. Such distributions are typical for sequential decay via the resonance or bound state of the subsystem; here the $n$-core $d$-state.

As seen the detectable large-distance distribution can be completely
different from the short-distance properties.  The $s$-wave dominance at large distance is necessary to produce
a halo configuration, while other partial waves may be present in the
short-distance bulk part. Thus, to learn about the
short-distance structure we have to exploit other observables or be
content with theoretically tracing the evolution of the wave function.

\paragraph{Conclusion.}

A new method is applied to investigate how nuclear halos are formed by clusterization from a mean-field background of bound nucleons. The existence of a slight, long-distance, $s$-wave tail is found to be sufficient for the formation of halos even when dominated by other spatially confined states. Strong evidence that Efimov states cannot exist in nuclei for one heavy and two light particles is provided. Finally, we show how large-distance observables can differ qualitatively from bulk properties at small distances, and how the evolution of the wave function can be traced.

%
%

\begin{acknowledgments}
This work was funded by the Danish Council for Independent Research DFF Natural Science and the DFF Sapere Aude program. This work has been partially supported by the Spanish Ministerio de Economia y Competitividad under Project FIS2014-51971-P. 
\end{acknowledgments}


\begin{thebibliography}{00}

  
\bibitem{vau72} D. Vautherin and D.M. Brink, Phys. Rev. C \textbf{5} (1972) 626.

\bibitem{dob16} J. Dobaczewski, J. Phys. G: Nucl. Part. Phys. {\bf 43} (2016) 04LT01. 

\bibitem{nik11} T. Niksi\'{c}, D. Vretenar, and P. Ring, Prog. Part. Nucl. Phys. {\bf 66} (2011) 519.

\bibitem{vau73} D. Vautherin, Phys. Rev. C \textbf{7} (1973) 296.

\bibitem{ben03} M. Bender, P.-H. Heenen, and P.-G. Reinhard, Rev. Mod. Phys. {\bf 75} (2003) 121.

\bibitem{bar13} B.R. Barrett, P. Navr\'{a}til, J.P. Vary, Prog. Part. Nucl. Phys. {\bf 69} (2013) 131.

\bibitem{epe09} E. Epelbaum, H.-W. Hammer, and U.-G. Meissner, Rev. Mod. Phys. {\bf 81} (2009) 1773. 

\bibitem{car15} J. Carlson, S. Gandolfi, F. Pederiva, S. C. Pieper, R. Schiavilla, K. E. Schmidt, and R. B. Wiringa, Rev. Mod Phys. \textbf{87} (2015) 1067.

\bibitem{her16} H. Hergert, S.K. Bogner, T.D. Morris, A. Schwenk, and K. Tsukiyama, Phys. Rep. \textbf{621} (2016) 165.

\bibitem{ono92} A. Ono, H. Horiuchi, T. Maruyama, A. Ohnishi, Prog. Theor. Phys. {\bf 87} (1992) 1185.

\bibitem{fel97} H. Feldmeier and J. Schnack, Prog. Part. Nucl. Phys. {\bf 39} (1997) 393.

\bibitem{hag10} G. Hagen, T. Papenbrock, D.J. Dean, M. Hjorth-Jensen, Phys. Rev. C {\bf 82} (2010) 034330.

\bibitem{tan78} Y.C. Tang and M. LeMere, Phys. Rep. {\bf 47} (1978) 167.
  
\bibitem{lei13} W. Leidemann and G. Orlandini, Prog. Part. Nucl. Phys. {\bf 68} (2013) 158.

\bibitem{hag14} G. Hagen, T. Papenbrock, M. Hjorth-Jensen, and D.J. Dean, Rep. Prog. Phys. {\bf 77} (2014) 096302.
                
\bibitem{lie16} S. Liebig, U.-G. Meissner, and A. Nogga, Eur. Phys. J. A {\bf 52} (2016) 103.

\bibitem{jen00} A.S. Jensen and K. Riisager, Phys. Lett. B \textbf{480} (2000) 39.
 
\bibitem{fre12} T. Frederico, A. Delfino, Lauro Tomio, M.T. Yamashita, Prog. Part. Nucl. Phys. {\bf 67} (2012) 939.

\bibitem{erl12} J. Erler, N. Birge, M. Kortelainen, W. Nazarewicz, E. Olsen, A. M. Perhac, and M. Stoitsov, Nature {\bf 486} (2012) 509.

\bibitem{nie01} E. Nielsen, D.V. Fedorov, A.S. Jensen, and E. Garrido, Phys. Rep. \textbf{347} (2001) 373.
  
\bibitem{fed93} D.V. Fedorov and A.S. Jensen, Phys. Rev. Lett. \textbf{71} (1993) 4103.
  
\bibitem{jen03} A.S. Jensen and D.V. Fedorov, Europhys. Lett. \textbf{62} (2003) 336.

\bibitem{gar06} E. Garrido, D. V. Fedorov, and A. S. Jensen, Phys. Rev. Lett. \textbf{96} (2006) 112501.

\bibitem{efi70} Efimov V., Phys. Lett. B \textbf{33} (1970) 563.

\bibitem{maz97} I. Mazumdar and V. S. Bhasin, Phys. Rev. C \textbf{56} (1997) R5(R).

\bibitem{maz00} I. Mazumdar, V. Arora, and V. S. Bhasin, Phys. Rev. C \textbf{61} (2000) 051303(R).

\bibitem{hag13} G. Hagen, P. Hagen, H.-W. Hammer, and L. Platter, Phys. Rev. Lett. \textbf{111} (2013) 132501.

\bibitem{hov17} D. Hove, E. Garrido, P. Sarriguren, D.V. Fedorov, H.O.U. Fynbo, A.S. Jensen, and N.T. Zinner, nucl-th/1702.01599, to be published.

\bibitem{gar99}  E. Garrido,  D.V. Fedorov, and A.S. Jensen, Nucl. Phys. A \textbf{650} (1999) 247.

\bibitem{cha98} E. Chabanat, P. Bonche, P. Haensel, J. Meyer, and F. Schaeffer, Nucl. Phys. A \textbf{635} (1998) 231. 
    
\bibitem{mad05} M. Bhattacharya and G. Gangopadhyay, Phys. Rev. C \textbf{72} (2005) 044318.

\bibitem{rii92}  K. Riisager,  A.S. Jensen and P. M{\o}ller,  Nucl. Phys. A \textbf{548} (1992) 393.

\bibitem{bra06} E. Braaten and H.-W. Hammer, Phys. Rep. {\bf 428} (2006) 259.

\bibitem{gar07} E. Garrido, D.V. Fedorov, H.O.U. Fynbo, and A.S. Jensen, Nucl. Phys. A {\bf 781} (2007) 387.

\bibitem{fyn09} H.O.U. Fynbo, R. \'{A}lvarez-Rodr\'{\i}guez, A.S. Jensen, O.S. Kirsebom, D.V. Fedorov, and E. Garrido, Phys. Rev. C \textbf{79} (2009) 054009.



\end{thebibliography}
\end{document}